# Growth kinetics and atomistic mechanisms of native oxidation of ZrS$_x$Se$_{2-x}$ and MoS$_2$ crystals

*Seong Soon Jo[1], Akshay Singh[2], Liqiu Yang[3], Subodh C. Tiwari[3], Sungwook Hong[4], Aravind Krishnamoorthy[3], Maria Gabriela Sales[5], Sean M. Oliver[7,8], Joshua Fox[6], Randal L. Cavalero[6], David W. Snyder[6], Patrick M. Vora[7,8], Stephen J. McDonnell[5], Priya Vashishta[3], Rajiv K. Kalia[3], Aiichiro Nakano[3], Rafael Jaramillo[1,\**

1. Department of Materials Science and Engineering, Massachusetts Institute of Technology, Cambridge, MA 02139, USA

2. Department of Physics, Indian Institute of Science, Bengaluru, Karnataka, 560012, India

3. Collaboratory for Advanced Computing and Simulation, University of Southern California, Los Angeles, CA 90089, USA

4. Department of Physics and Engineering, California State University, Bakersfield, Bakersfield, CA 93311, USA

5. Department of Materials Science and Engineering, University of Virginia, Charlottesville, VA 22904, USA

6. Electronic Materials and Devices Department, Applied Research Laboratory and 2-Dimensional Crystal Consortium, Materials Research Institute, Pennsylvania State University, University Park, PA 16802, USA

7. Department of Physics and Astronomy, George Mason University, Fairfax, VA 22030, USA

8. Quantum Materials Center, George Mason University, Fairfax, VA 22030, USA

* Corresponding Author

E-mail: rjaramil@mit.edu

**Abstract**

A thorough understanding of native oxides is essential for designing semiconductor devices. Here we report a study of the rate and mechanisms of spontaneous oxidation of bulk single crystals of $ZrS_xSe_{2-x}$ alloys and $MoS_2$. $ZrS_xSe_{2-x}$ alloys oxidize rapidly, and the oxidation rate increases with Se content. Oxidation of basal surfaces is initiated by favorable $O_2$ adsorption and proceeds by a mechanism of Zr-O bond switching, that collapses the van der Waals gaps, and is facilitated by progressive redox transitions of the chalcogen. The rate-limiting process is the formation and out-diffusion of $SO_2$. In contrast, $MoS_2$ basal surfaces are stable due to unfavorable oxygen adsorption. Our results provide insight and quantitative guidance for designing and processing semiconductor devices based on $ZrS_xSe_{2-x}$ and $MoS_2$, and identify the atomistic-scale mechanisms of bonding and phase transformations in layered materials with competing anions.

A thorough understanding of the processing and properties of native oxides is essential for designing semiconductor devices. This is no less true for nanomaterials than it is for legacy semiconductors such as silicon, for which control and understanding of the native oxide was a seminal achievement of 20$^{th}$ century materials science. Layered transition metal dichalcogenides (TMDs) nominally have inert, fully-passivated surfaces, but it is well-known that this is an oversimplification and that many TMDs oxidize readily. As interest grows in applications of TMDs in microelectronics, photonics, and optoelectronics, so too does the importance of developing predictive materials processing methods that take into account the native oxide.[1–6]

The stability of TMDs in ambient conditions has been studied for decades.[7–23] It is known that stability against oxidation increases as the chalcogen element changes from Te, to Se, to S, and that TMDs including group-4 transition metals are particularly unstable in air.[7,10–12] Monolayers are less stable than few-layer and bulk crystals, but the basal surfaces of monolayer $MoS_2$ and $MoSe_2$ are stable against chemisorption in ambient conditions, and oxidation of few-layer $WS_2$ is self-limiting.[7,8,13–15,21,24,25] Oxidation preferentially begins at defects such as chalcogen vacancies and grain boundaries, can be accelerated by illumination, and may be facilitated by organic absorbates.[7,9,14–16,24,26] The effects of humidity on oxidation have also been studied.[17–20] For instance, the oxidation rate of pulverized $MoS_2$ at elevated temperature has been found to depend on relative humidity and particle size.[20] It has been shown that $H_2O$ is a milder oxidizing agent than $O_2$ for oxidation of epitaxially-grown monolayer $MoS_2$ at elevated temperature.[21] These results suggest that $H_2O$ may be a catalyst for spontaneous oxidation in humid air. TMD oxidation has also been studied in harsh environments typical of device processing, such as exposure to ozone or etchant.[19,22,27–29]

Here we report on the growth kinetics and atomic mechanisms of native oxidation in ambient conditions of freshly-cleaved surfaces of bulk single crystals of $ZrS_xSe_{2-x}$ alloys and $MoS_2$. We find that the Zr-based TMDs oxidize readily, with the rate increasing with Se content. Meanwhile, $MoS_2$ seems to oxidize not at all in air on a laboratory time scale. We quantify the oxide growth rate with sub-nm resolution using spectroscopic ellipsometry (SE).[30] SE represents an improvement in thickness resolution and measurement time over other methods used to observed TMD oxidation, such as atomic force microscopy, optical microscopy, and Raman spectroscopy.[8–11] Our results are unique in quantifying oxidation of a TMD alloy series, and may be particularly useful for processing electronic devices from Zr-based TMD nanomaterials.[31]

Further, we use first principles-validated reactive molecular dynamics (RMD) simulations to illustrate the atomistic mechanism of oxidation of $ZrS_xSe_{2-x}$, and to understand the stability of $MoS_2$. We find that oxygen favorably chemisorbs on the $ZrS_2$ surface, and then moves readily into the crystal lattice by a mechanism of Zr-O bond switching, facilitated by successive redox transitions at the S sites. At short times, the interaction between O and S is indirect, mediated by the Zr atoms. The rate-limiting step for oxide growth is the formation and out-diffusion of $SO_2$. In contrast, we find the oxidation of $MoS_2$ is kinetically inhibited by a large energetic barrier to $O_2$ dissociative adsorption, even though chemisorbed oxygen is thermodynamically preferred.[25] These results are enabled by the development of a force field that can describe Zr-

Mo-S-O interactions, an advance over previous work using first principles-informed RMD simulations based on ReaxFF reactive force fields to study the Mo-S-O system.[32]

In **Fig. 1** we present the imaginary component of the effective dielectric constant ($\varepsilon_2^{\text{eff}}$) measured on single-crystal ZrS$_x$Se$_{2-x}$ alloys as a function of time after cleaving, measured by SE. The effective dielectric constant accurately represents the material only for a uniform, semi-infinite sample with an intimate and smooth sample-air interface, hence the qualifier "effective" (see Supplementary Information, SI). The time-dependence of the data is a straightforward clue that the actual sample structure is more complicated, and changes in time as the samples are exposed to air. Comparing the results for ZrSe$_2$ (**Fig. 1a**) to the results for ZrS$_2$ (**Fig. 1f**), we see that the sulfide is more air-stable.[12] The peaks in the $\varepsilon_2^{\text{eff}}$ spectra derive from the lowest-lying direct, allowed optical transitions and excitons ($X_\Gamma$, $X_L$, $X_A$), which blue-shift with sulfur content across the ZrS$_x$Se$_{2-x}$ series.[33]

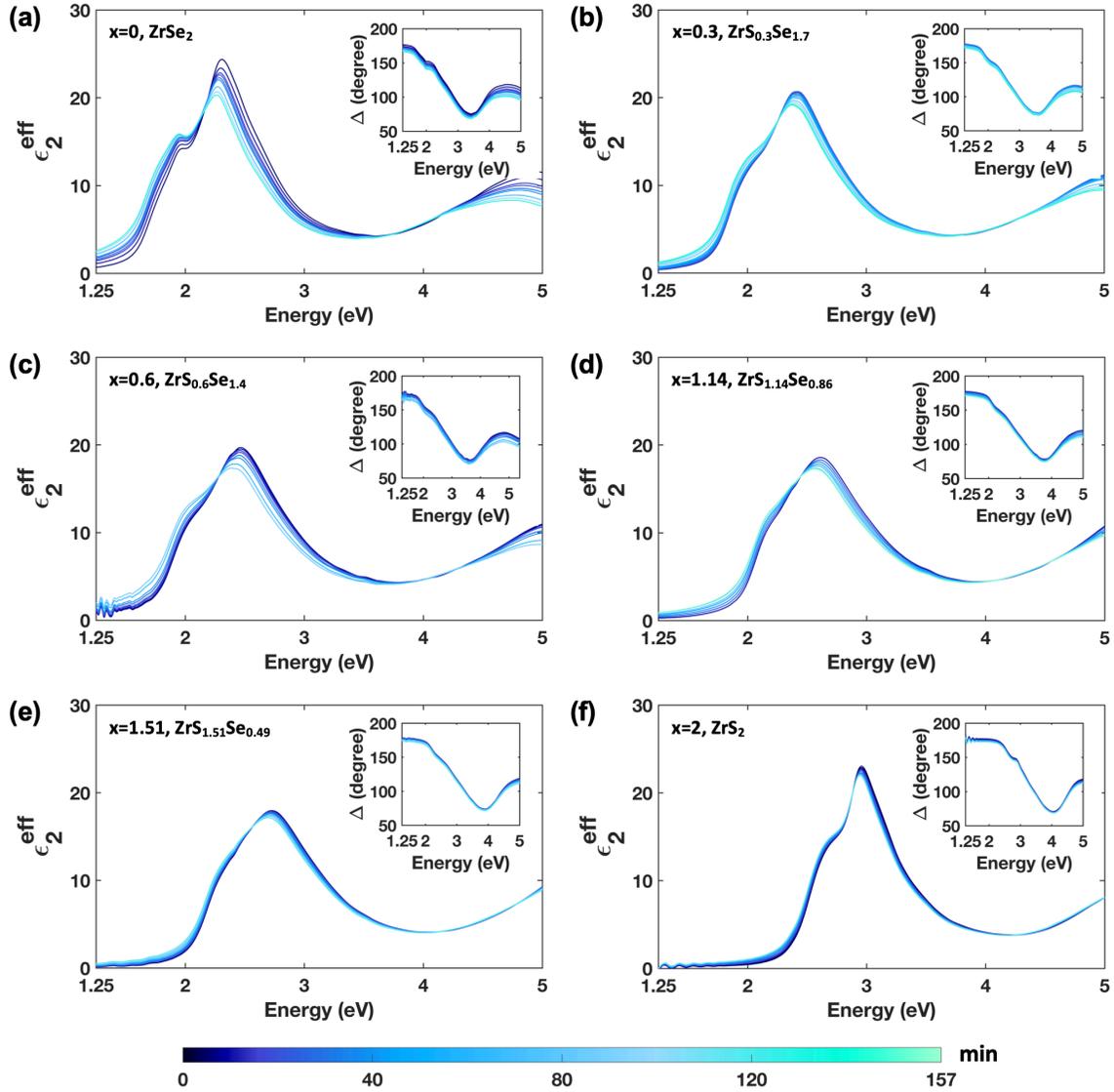

**Figure 1:** Imaginary part of the effective dielectric constant ($\varepsilon_2^{\text{eff}}$) measured by SE, as a function of time in ambient conditions after exposing a fresh surface, for six compositions: **(a)** $ZrSe_2$, **(b)** $ZrS_{0.3}Se_{1.7}$, **(c)** $ZrS_{0.6}Se_{1.4}$, **(d)** $ZrS_{1.14}Se_{0.86}$, **(e)** $ZrS_{1.51}Se_{0.49}$, **(f)** $ZrS_2$. The insets show the SE amplitude ($\Delta$) for the same data series. The color bar at the bottom indicates the exposure time after cleaving.

The interpretation of the $\varepsilon^{\text{eff}}$ data presented in **Fig. 1** deserves some discussion. In the presence of a native oxide, $\varepsilon^{\text{eff}}$ is not an accurate estimate of the actual dielectric response of the underlying crystal. In this case, the actual dielectric constant must be determined through modeling and regression (**Sec. S2**). We assume that the dielectric properties of the oxide and the underlying crystal are unchanging in time, and that the time dependence of the data derives from increasing oxide thickness. We assume that the oxide is a dielectric with refractive index spectra equal to that of ZrO$_2$ (using reference data) or MoO$_3$.[34] We use this model to analyze the experimental data, leading to the thickness-time results presented below. To aid interpretation of the $\varepsilon^{\text{eff}}$ data presented in **Fig. 1**, we simulate $\varepsilon_2^{\text{eff}}$ for different oxide thickness, all else held constant (**Fig. S2**). We see that $\varepsilon_2^{\text{eff}}$ overestimates (underestimates) the extinction coefficient ($\kappa = -\sqrt{\varepsilon_2}$) well below (well above) the exciton resonances. We also see that $\varepsilon_2^{\text{eff}}$ suggests a red-shift of the exciton resonances with increasing oxide thickness. This is a subtle effect of light interacting with a multi-layer sample with absorption resonances; it does not imply a change to the exciton spectrum structure of the underlying crystal. The simulated $\varepsilon_2^{\text{eff}}$ spectra also indicate the ability of SE to resolve oxide thickness down to 2 Å, below which the spectra cease changing. This is comparable to a metal-oxygen bond length.

In **Fig. 2** we show SE time-series data recorded after cleaving a fresh surface of MoS$_2$. In contrast to the results for ZrS$_x$Se$_{2-x}$, the results for MoS$_2$ show no apparent change, even after many days of exposure. This suggests that a pristine MoS$_2$ surface does not spontaneously oxidize in air on laboratory time scales.

If freshly-cleaved MoS$_2$ does maintain a pristine surface in air, then the effective dielectric constant is an accurate estimate of the optical properties of the material. In **Fig. 2a** we show the complex dielectric constant ($\varepsilon_1 - i\varepsilon_2$), and in **Fig. 2b** we show the complex refractive index ($n - i\kappa$). The optical transitions characteristic of MoS$_2$ excitons A ($\approx$ 1.8 eV), B ($\approx$ 2.1 eV), C ($\approx$

2.7 eV) and D ($\approx$ 3.1 eV) are clearly observed in the loss spectra ($\varepsilon_2$ and $\kappa$). We note the small loss coefficient, $\kappa < 0.05$, for photon energy near and below the indirect band gap of 1.29 eV. This value is lower than our previously-reported value of $\kappa \approx 0.2$, which we determined from SE measurements on geological crystals from the same batch but without cleaving.[5] This illustrates the sensitivity of SE to surface conditions, including roughness. The value of the loss coefficient that we measure here is comparable to what we measure on waveguide-integrated synthetic $MoS_2$ thin films ($\kappa \approx 0.04 - 0.1$, unpublished), and approaches our theoretical predictions for ideal, bulk $MoS_2$ ($\kappa \approx 0.01 - 0.02$).[4,5] These value bode well for refractive, low-loss applications of $MoS_2$ in integrated photonics.[35]

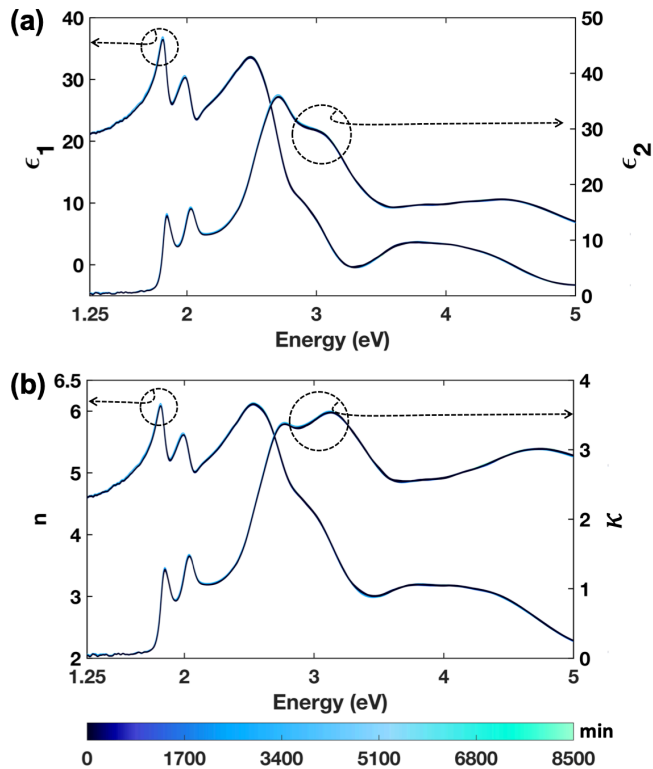

**Figure 2:** Complex dielectric constant **(a)** and refractive index **(b)** of $MoS_2$ determined by SE measurements on a freshly-cleaved surface. The color series indicates the time elapsed between cleaving the surface and taking the SE data. The spectra recorded at times between 6

and 8,498 min overlap and are barely distinguishable. This time-invariance suggests that the freshly-exposed MoS$_2$ surface is stable in air on laboratory time scales.

In **Fig. 3a-b** we plot the time-dependence of the growth of native oxide on ZrS$_x$Se$_{2-x}$ and MoS$_2$. The oxidation of Se-rich ZrS$_x$Se$_{2-x}$ starts immediately after cleaving and the oxide grows approximately logarithmically in time, showing no sign of stopping within the 160 min observation window. S-rich samples oxidize more slowly; for ZrS$_2$ the oxidation process appears to stop above $\approx$ 1 nm of oxide thickness. In the SI we discuss the applicability of established kinetic models to interpret our results.

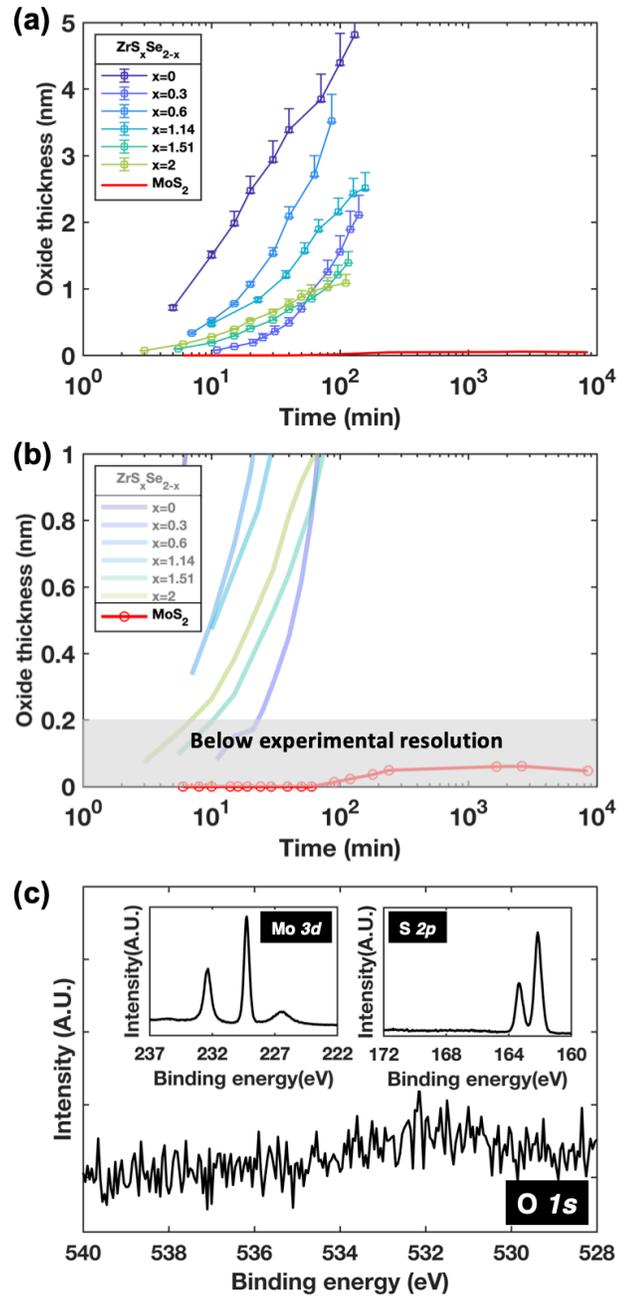

**Figure 3:** Kinetics of native oxide formation on freshly-cleaved $ZrS_xSe_{2-x}$ and $MoS_2$. **(a)** Oxide thickness *vs.* exposure time for $ZrS_xSe_{2-x}$ alloys; from blue to light green, x=0, x=0.3, x=0.6, x=1.14, x=1.51, x=2. The error bars represent systematic errors in the model and nonlinear regression procedure. It reflects $\pm$ 5% error in initial guess of oxide thickness and 10% error in surface roughness of oxide. **(b)** Plot of oxide thickness *vs.* exposure time for MoS$_2$. The model-based data analysis infers an oxide thickness that is below the experimental resolution of $\approx$ 2 Å, which is indicated by the grey region. **(c)** XPS O-1s peak and (inset) Mo-3d peak and S-2p peak for a MoS$_2$ crystal stored in laboratory ambient conditions for a year after cleaving. No oxygen signal is detected.

We find that MoS$_2$ barely oxidizes in air after cleaving, if at all. Our SE modeling and data analysis is consistent with an oxide thickness of $\approx$ 1 Å forming within 5 hrs after cleaving, and remaining stable after 6 days on air exposure, as we show in **Fig. 3b**. However, this result is below the experimental sensitivity, and therefore our data is also consistent with a stable and pristine MoS$_2$ surface. To distinguish between these two scenarios – incipient oxidation, or a stable and pristine MoS$_2$ surface – we turn to XPS. As we show in **Fig. 3c**, the O 1s spectrum shows no detectable Mo-O bonding (530.2 eV for MoO$_3$), and the S 2p spectrum shows no S-O bonding (167 eV for SO$_2$), indicating that the MoS$_2$ surface is not oxidized. These XPS results were obtained from a geological MoS$_2$ crystal that was mechanically cleaved and then exposed to ambient conditions for a year. Both SE and XPS show that the surface of MoS$_2$ is stable and non-reactive in typical laboratory conditions, as has been reported previously.[10,25]

To understand the rapid oxidation of Zr-based TMDs compared to MoS$_2$, we perform reactive molecular dynamics (RMD) simulations of the oxidation of MoS$_2$ and ZrS$_2$ slabs. The simulation volume consists of four TMD layers. While this is far thinner than the bulk crystals that we study experimentally, it is sufficient to study the key atomic-scale mechanisms, and still requires substantial computational resources. We simulate oxidation in an oxygen partial pressure of 1.16 kbar and at 800 K, to accelerate the process; as a result, the timescales of oxidation observed in simulation are not expected to match those observed in experiment. In **Fig. 4** we show the initial simulation setup (at time $t$ = 0 ns) and the final snapshot ($t$ = 2.5 ns) for MoS$_2$ and ZrS$_2$ oxidation. The simulation box is periodic in the ⟨100⟩ and ⟨010⟩ directions (*i.e.*, in-plane). The simulations show that ZrS$_2$ oxidizes extensively. To quantify the extent of oxidation, we show the total numbers of M-S, M-O, O-O, and S-O bonds (M = Mo or Zr). The increase in Mo-O with oxidation time is negligible compared to the rapid increase of Zr-O bonds. Concurrently, we observe a notable decrease in O-O bonds in the ZrS$_2$ simulation, suggesting scission of O-O bonds followed by formation of Zr-O bonds. The bond counting data in **Fig. 4f** suggest that ZrS$_2$ oxidation progresses in two stages. During the first 0.1 ns, the number of Zr-O bonds rises rapidly, followed by much slower increase at longer times. The rapid initial reaction may be explained by energetically-favored O$_2$ adsorption on the ZrS$_2$ surface.[36] The rapid increase in the number of Zr-O bonds is not accompanied by a similarly rapid decrease in the number of Zr-S bonds, as we discuss below. The subsequent, slower process of Zr-O bond formation represents the growth of an amorphous oxy-sulfide that largely eliminates the van der Waals gaps. In contrast, we observe almost no decrease of O-O bonds in the MoS$_2$ simulation. This suggests that MoS$_2$ remains mostly inert and stable in the simulated environment. We also find that the presence of a grain boundary substantially accelerates oxidation of ZrS$_2$, as expected; we present and discuss these results in **Sec. S9**.

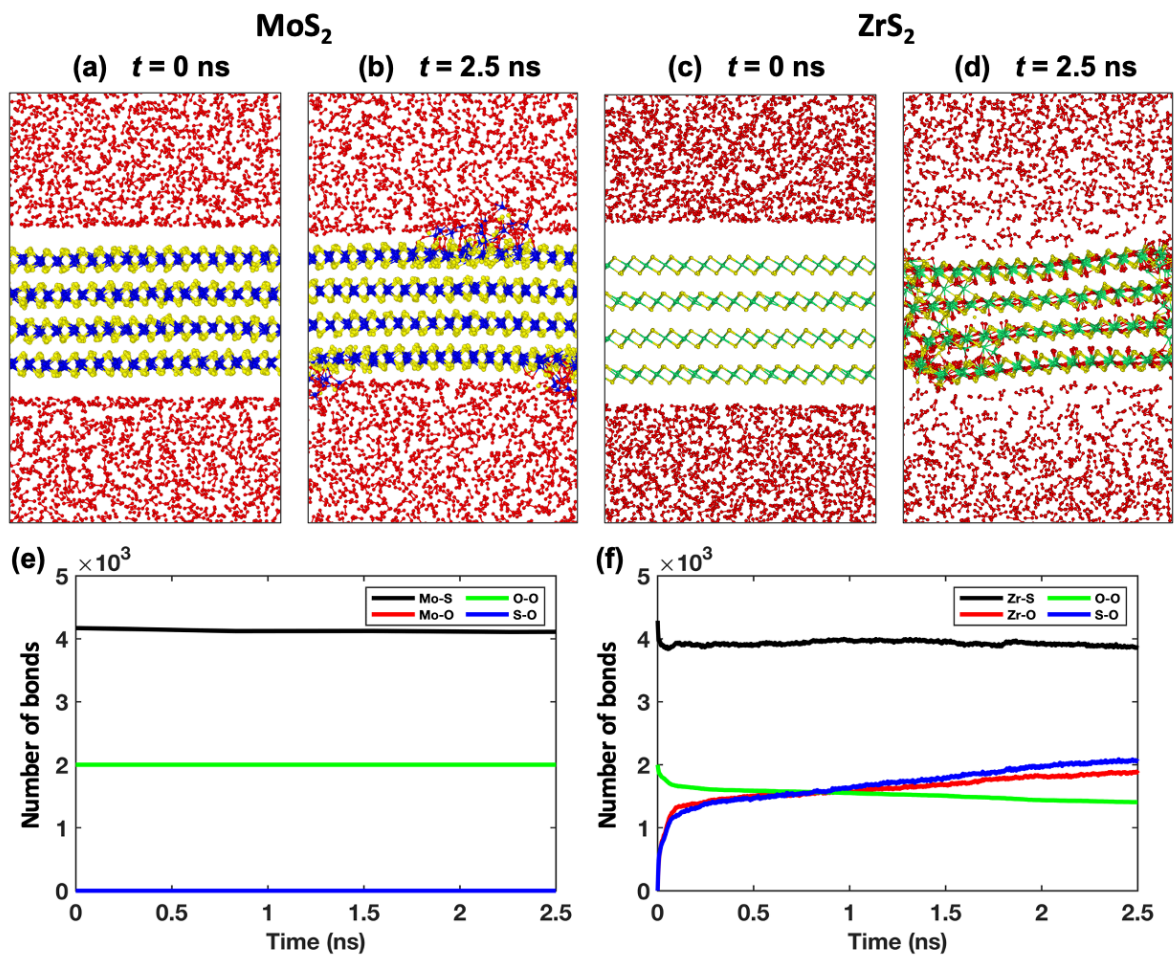

**Figure 4:** RMD simulations of MoS$_2$ and ZrS$_2$ oxidation. Snapshots of **(a-b)** MoS$_2$ oxidation and **(c-d)** ZrS$_2$ oxidation at time $t$ = 0 and 2.5 ns, respectively. Spheres represent individual atoms: blue (Mo), green (Zr), yellow (S) and red (O). Time evolution of corresponding bond count for **(e)** MoS$_2$ and **(f)** ZrS$_2$ oxidation.

We analyze the process of ZrS$_2$ oxidation in detail by considering atomic trajectories. We find that oxidation proceeds in six steps: (1) adsorption of O$_2$ molecules on the ZrS$_2$ surface; (2) dissociation of O$_2$ on ZrS$_2$ surface, and formation of Zr-O bonds; (3) Zr pushed into van der Waals gaps between TMD layers, due to formation of new Zr-O bonds; (4) formation of inter-layer Zr-S and Zr-O bonds, resulting in breakdown of the layered crystal structure; (5) oxygen transport laterally within and vertically between layers by a mechanism of Zr-O bond switching; (6) formation and out-diffusion of $SO_2$. In **Fig. 5a-b** we show the cross-section and top-down view of the simulation at $t$ = 2 ps, illustrating steps 1-2. We indicate $O_2$ adsorption sites with black rectangles. We find that the adsorption of $O_2$ on the $ZrS_2$ basal surface is energetically-favorable, with adsorption energy $E_{ads}$ = -0.006 eV per $O_2$ molecule, which we calculate by density function theory (DFT) simulations of a $ZrS_2$ surface-$O_2$ molecule system. In contrast, we find no stable, relaxed energetic minimum for a $MoS_2$ surface-$O_2$ molecule system, indicating that the adsorption energy is positive. Constrained MD simulations paint a similar picture, that $O_2$ adsorption is favorable on $ZrS_2$ but unfavorable on $MoS_2$ (see **Fig. S5**). After rapid adsorption of $O_2$, we observe O-O bond breaking and the formation of Zr-O bonds. We also observe 2- and 3-coordinated oxygen atoms on the $ZrS_2$ surface after O-O bond breakage, corresponding to 2-coordinated bridge oxygen, and 3-coordinated oxygen that are seeds for zirconia growth. A similar process was proposed for the oxidation of monolayer $ZrS_2$.[23]

Following the initial formation of Zr-O bonds, Zr moves from the top layer into the interlayer region. The protrusion of Zr atoms results in the collapse of the van der Waals gap between $ZrS_2$ layers by the formation of new, interlayer Zr-S and Zr-O bonds, resulting in amorphous, non-layered material. This is illustrated by a simulation view at $t$ = 2.5 ns, presented in **Fig. 5c-d**. The close-up in **Fig. 5d** highlights the inter-layer Zr atoms and bonds. The expansion of these disordered regions into the van der Waals gaps provides pathways for oxygen transport further into the $ZrS_2$ crystal. We find that oxygen transport proceeds by a process of Zr-O bond switching, illustrated step-by-step in **Fig. 5e**, in which Zr and O atoms are numbered and S atoms are omitted for clarity. The coordination of oxygen atom O1 first changes from 3 to 2 by the breakage of the Zr1-O1 bond. The subsequent creation of a new Zr4-O1 bond restores the 3-fold coordination. A similar process continues when the Zr2-O1 bond is broken and a new Zr5-O1 bond is formed. Oxygen thus traverses the system by successive breaking and forming Zr-O bonds. This bond-switching mechanism for oxygen transport is akin to Grotthuss mechanism which plays essential roles in various processes in oxides.[37–39] It is interesting that this process is not correlated with a decrease in the number of Zr-S bonds. This suggests that S atoms act as spectators during these initial steps of Zr-O bond formation, amorphization, and oxygen transport, and that charge neutrality is maintained by successive redox transitions at the S sites. The intermediate phase is therefore a zirconium oxy-sulfide, with mixed-valence sulfur ions. The thermodynamic end-products of $ZrS_2$ oxidation are $ZrO_2$ and $SO_2$; therefore, the oxidation state of sulfur goes from $S^{2-}$ to $S^{4+}$. It is interesting that the oxidation of sulfur is mediated initially through the Zr site. These processes are also observed in first-principles QMD simulations (**Figs. S5 and S6**), thus validating these atomistic mechanisms of $ZrS_2$ oxidation.

The last step in the oxidation of $ZrS_2$ is the formation and out-diffusion of $SO_2$. Our simulation at 800 K does not extend for long enough time to observe this process. Simulations performed at a much higher temperature of 1500 K do observe the formation and out-diffusion of $SO_2$ molecules (**Fig. S7**).

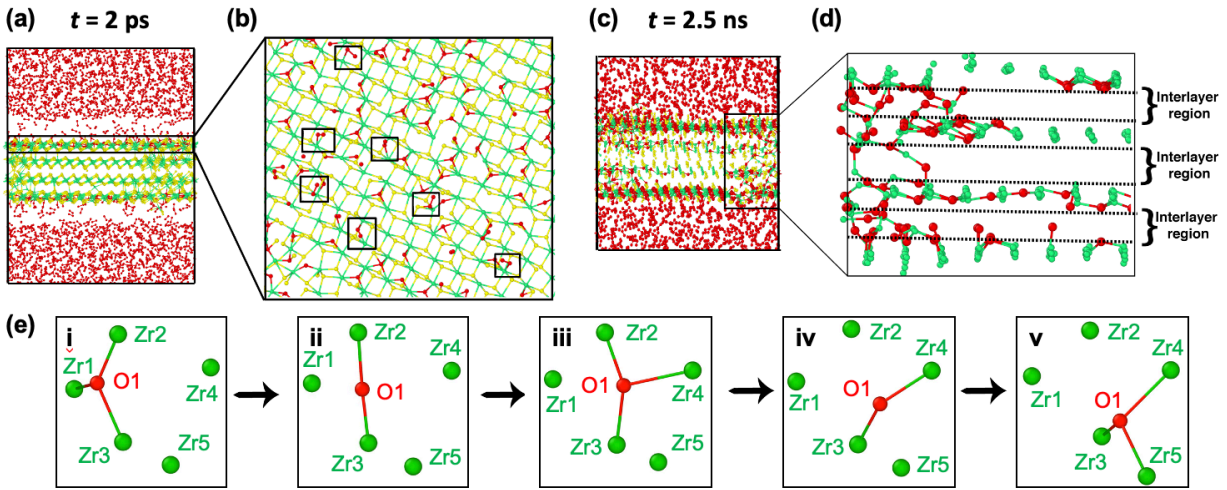

**Figure 5.** Atomistic mechanisms of ZrS$_2$ oxidation. Cross section **(a)** and top-down **(b)** views at $t$ = 2 ps, showing O$_2$ adsorption and Zr-O bond formation that involves both 2- and 3-fold coordinated oxygen atoms (enclosed by black rectangles). Cross section **(c)** and close-up **(d)** view at $t$ = 2.5 ns, showing the formation of an amorphous matrix and the closing of the van der Waals gaps; S atoms are not shown in **(d)** for clarity. (**e**) Oxygen transport mechanism of Zr-O bond switching, including oxygen switching between 2- and 3-fold coordination: (i-ii) breakage of Zr1-O1 bond and decrease in coordination number (3→2); (ii-iii) formation of Zr4-O1 bond and increase in coordination number (2→3); (iii-iv) breakage of Zr2-O1 bond and reduced coordination number (3→2); (iv-5) formation of Zr5-O1 bond and coordination number recovery (2→3).

Our results provide quantitative guidance for designing and processing semiconductor devices based on ZrS$_x$Se$_{2-x}$ and MoS$_2$, and identify the atomistic-scale mechanisms of bonding and phase transformations in layered materials with competing anions. Direct, atomic-scale experimental testing of the mechanisms illustrated here by MD may require extensive use of methods of chemical spectroscopy with Å-level depth resolution, including variable-angle photoelectron spectroscopy, and possibly the development of tools to resolve bond forming and breaking on ultra-fast time scales. It will be interesting to extend our methods to study the oxidation of thin films produced by various methods, especially including wafer-scale processes that will be essential for device fabrication, and in different conditions.[4] This is especially pertinent if native oxides are discovered to have beneficial passivating or functional properties, as they do for legacy semiconductors.[31]

## Acknowledgements

This work was supported by an Office of Naval Research MURI through grant #N00014-17-1-2661. This work was supported by a start-up grant from the Indian Institute of Science. We acknowledge the use of facilities and instrumentation supported by NSF through the Massachusetts Institute of Technology (MIT) Materials Research Science and Engineering Center DMR - 1419807. This material is based upon work sponsored in part by the U.S. Army Research Office through the Institute for Soldier Nanotechnologies, under contract number W911NF-13-D-0001. The work at USC was supported as part of the Computational Materials Sciences Program funded by the U.S. Department of Energy, Office of Science, Basic Energy Sciences, under Award Number DE-SC0014607. We would also like to thank USC High Performance Computing Center for provided computing resources. The work was financially supported by the National Science Foundation (NSF) through the Pennsylvania State University 2D Crystal Consortium – Materials Innovation Platform (2DCC-MIP) under NSF cooperative agreement DMR-1539916. P.M. Vora and S.M. Oliver acknowledge support from the George Mason University Quantum Materials Center, Presidential Scholars Program, and from the NSF through Grant No. 1847782. We acknowledge assistance from the Department of Mineral Sciences, Smithsonian Institution. M.G.S. acknowledges support through the William L. Ballard Jr. Endowed Graduate Fellowship.

**Supporting Information Available.** The following files are available for free of charge *via* the Internet at http://pubs.acs.org :

S1. Experimental Methods; S2. Optical modelling – oxide and surface roughness; S3. Theory and simulation methods; S4. ReaxFF reactive force field development for the Zr/S interactions; S5. $O_2$ adsorption energy on $MoS_2$ and $ZrS_2$ surfaces by quantum-mechanical calculation; S6. Validation of $ZrS_2$ force field by quantum molecular dynamics simulation; S7. $ZrS_2$ oxidation at 1500 K; S8. Kinetic models for oxide growth; S9. Grain boundary oxidation in $ZrS_2$; S10. Surface morphology (PDF)

A ReaxFF parameter set for Mo/Zr/S/O interactions (TXT)

AUTHOR INFORMATION

Corresponding Author

E-mail: rjaramil@mit.edu

**Table of Contents Graphic**

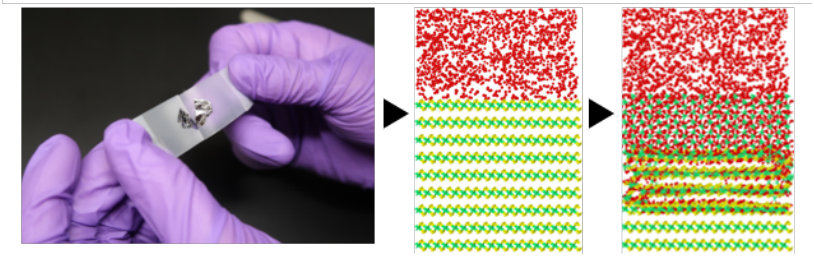